\documentclass[reprint,amsmath,amssymb,aip,jcp,floatfix,eqsecnum]{revtex4-1}

\usepackage{graphicx}
\usepackage{dcolumn}
\usepackage{bm}

 \usepackage{color}

\usepackage{amssymb}
\usepackage{amsmath}
\usepackage{graphicx}
\usepackage{amsbsy}

\usepackage{enumerate}

\newcommand\beq{\begin{equation}}
\newcommand\eeq{\end{equation}}
\newcommand\beqa{\begin{eqnarray}}
\newcommand\eeqa{\end{eqnarray}}
\newcommand{\nn}{\nonumber\\}

\newcommand{\id}{\text{id}}
\newcommand{\ex}{\text{ex}}
\newcommand{\mm}{s}

\newcommand{\py}{\text{PY}}
\newcommand{\hnc}{\text{HNC}}

\def\bal#1\eal{\begin{align}#1\end{align}}

\begin{document}



\title{Fourth virial coefficient of additive hard-sphere mixtures
in the Percus--Yevick and hypernetted-chain approximations}


\author{Elena Beltr\'an-Heredia}
\author{Andr\'es Santos}
\email{andres@unex.es}
\homepage{http://www.unex.es/eweb/fisteor/andres/}
\affiliation{Departamento de F\'{\i}sica, Universidad de
Extremadura, Badajoz, E-06071, Spain}

\date{\today}

\begin{abstract}
The fourth virial coefficient  of additive hard-sphere mixtures, as predicted by the Percus--Yevick (PY) and hypernetted-chain (HNC) theories, is derived  via the compressibility, virial, and chemical-potential routes,  the outcomes being compared  with exact  results. Except in the case of the HNC compressibility route, the other five expressions exhibit a common structure involving the first three moments of the size distribution. In both theories the chemical-potential route is slightly better than the virial one and the best behavior is generally presented by the compressibility route. Moreover, the PY results with any of the three routes are more accurate than any of the HNC results.
\end{abstract}

\date{\today}


\maketitle
\section{Introduction}
\label{sec1}
As is well established, the hard-sphere (HS) model plays a paradigmatic role in statistical physics, both in and out of equilibrium.\cite{BH76,HM06,L01,M08,S13}
The importance of HS systems in equilibrium liquid state theory is incremented by the existence  of exact solutions of the Percus--Yevick (PY) integral equation theory\cite{HM06,PY58} both for pure\cite{W63,T63,W64} and multicomponent\cite{L64} HS fluids. In contrast, other integral equations, like the hypernetted-chain (HNC) one,\cite{M58,vLGB59,M60} need to be solved numerically, even for HS systems.

A consequence of dealing with solutions of approximate integral equations (like PY and HNC) is that a common pair correlation function, when plugged into the so-called thermodynamic routes, gives rise to different equations of state.\cite{HM06,S13b} This inconsistency problem is aggravated in the case of HS mixtures since the corresponding pressure depends not only on density but also on the size distribution.

The exact equation of state of HS mono- or polydisperse fluids is not analytically known, and so one has to rely on computer simulation results to assess the merits and drawbacks of approximate theories. An exception is provided by the low-density regime, in which case the equation of state can be well represented by the first few virial coefficients.
The first four coefficients of pure HS fluids are analytically known\cite{CM04b,L05} and accurate  numerical evaluations of the 5th to 12th coefficients can be found in the literature.\cite{CM04a,LKM05,CM05,CM06,W13}
Much less information is available for HS mixtures, the results being usually restricted to the binary case. While the second and third virial coefficients are exactly known  for additive and nonadditive mixtures with any number of components,\cite{K55,KM75,H96,H96b,AEH99} the fourth to eighth coefficients have been numerically computed for binary mixtures at a number of size ratios and/or nonadditivities.\cite{SFG96,SFG96b,SFG97,EACG97,EAGB98,WSG98,SFG98,VM03,PCGS07,HML10}
Recently, analytical expressions for all but one of  the partial contributions to the fourth virial coefficient $B_4$ of additive binary mixtures have been derived\cite{LK09,U11} and a very accurate semi-empirical equation for the last contribution has been constructed.\cite{LK09}

The aim of this paper is to evaluate  $B_4$ (for additive HS mixtures) predicted by the PY and HNC approximations via different thermodynamic routes, and to compare them with the exact (analytical and semi-empirical) results. The two conventional routes in the case of HS systems are the compressibility and virial ones.\cite{BH76,HM06} While $B_4$ from the PY theory follows from the corresponding known equations of state\cite{L64,YSH98} in a straightforward way, we are not aware of a previous derivation of $B_4$ for polydisperse HS systems in the HNC theory via the virial and compressibility routes. To those conventional routes, we add the results derived from the chemical-potential route,\cite{S12b,SR13,RS14} usually not considered in the literature. As will be shown, the PY results with any of the three routes are more accurate than any of the HNC results. Moreover, within a given theory (PY or HNC), the best behavior is due to the compressibility route, the chemical-potential results being slightly better than the virial ones.

The organization of the rest of the paper is as follows. The background material, including exact results, are presented in Sec.\ \ref{sec2}. Next, the PY and HNC results are derived in Sec.\ \ref{sec3} and discussed in Sec.\ \ref{sec4}. The paper is closed with some concluding remarks in Sec.\ \ref{sec5}.

\section{Background}
\label{sec2}
\subsection{Thermodynamic routes}
Let us consider a multicomponent HS system made of $\mm$ components. The interaction between a particle of species $i$ and a particle of species $j$ is
\beq
\varphi_{ij}(r)=\begin{cases}
  \infty, &r<\sigma_{ij},\\
  0,&r>\sigma_{ij},
  \end{cases}
  \label{1a}
\eeq
where $\sigma_{ij}$ is the closest distance of separation for both particles. If we denote by $\sigma_i\equiv \sigma_{ii}$ and $\sigma_j\equiv \sigma_{jj}$ the diameters of particles of species $i$ and $j$, respectively, we say that the HS mixture is \emph{additive} if $\sigma_{ij}=\frac{1}{2}(\sigma_i+\sigma_j)$ for all pairs $(ij)$. Otherwise, the mixture is said to be \emph{nonadditive}. The thermodynamic  state of the mixture is characterized by the total number density $\rho=N/V$ (where $V$ and $N$ are the volume and the total number of particles, respectively) and the mole fractions $\{x_i\}$ (with the constraint $\sum_i x_i=1$).

In general, the knowledge of the set of radial distribution functions $g_{ij}(r;\rho)$ or, equivalently, the set of cavity functions
\beq
y_{ij}(r;\rho)=e^{\beta\varphi_{ij}(r)} g_{ij}(r;\rho),
\label{2a}
\eeq
where $\beta=1/k_BT$ is the inverse temperature, allows one to obtain the thermodynamic properties of the fluid through the so-called thermodynamic routes.\cite{HM06,S13b} In particular, the compressibility route reads
\begin{equation}
\chi_{T}^{-1}(\rho)\equiv\left(\frac{\partial\beta p}{\partial\rho}\right)_{T}=\sum_{i,j}\sqrt{x_ix_j}\left[\mathsf{I}+\mathsf{\hat{h}}(\rho)\right]_{ij}^{-1},
\label{eq:104}
\end{equation}
where $p$ is the pressure, $\mathsf{I}$ is the $\mm\times \mm$ unity matrix and the matrix $\mathsf{\hat{h}}$ is defined as
\begin{equation}
\hat{h}_{ij}(\rho)=4\pi\rho\sqrt{x_{i}x_{j}}\int_0^\infty{d}{r}\,r^2\left[g_{ij}(r;\rho)-1\right].
\label{hij}
\end{equation}
The associated compressibility factor $Z\equiv{\beta p}/{\rho}$ can be obtained as
\beq
Z^{(c)}(\rho)=\int_0^1 dt\, \chi_T^{-1}(\rho t).
\label{Zc}
\eeq
The superscript in $Z^{(c)}$ means that the compressibility route has been used.

Particularized to HS mixtures, the virial (or pressure) route is
\begin{equation}
Z^{(v)}(\rho)=1+\frac{2\pi}{3}\rho\sum_{i,j}x_{i}x_{j}\sigma_{ij}^{3}y_{ij}(\sigma_{ij};\rho).
\label{eq:101-1}
\end{equation}

Finally, we are here especially interested in the chemical-potential route, which has received little attention until recently.\cite{S13b,S12b,SR13} In the case of HS mixtures,\cite{SR13}
\bal
\beta\mu_{i}(\rho)&=\beta\mu_{i}^\id(\rho)+\beta\mu_{i}^\ex(\rho)\nn
&=\ln\left(\rho x_{i}\Lambda_{i}^{3}\right)+4\pi \rho\sum_{j}x_{j}\int_{0}^{\sigma_{ij}}{d}\sigma_{0j}\,\sigma_{0j}^{2}y_{0j}(\sigma_{0j};\rho),
\label{eq:102-1}
\eal
where $\Lambda_i={h}\sqrt{\beta/2\pi m_i}$ ($h$ and $m_i$ being the Planck constant and the mass of a particle of species $i$, respectively) is the thermal de Broglie wavelength. In Eq.\ \eqref{eq:102-1},  $\{y_{0j}(r)\}$ are the set of  cavity functions associated with an ``impurity'' particle that interacts with a particle of species $j$ via a HS potential characterized by the distance $\sigma_{0j}$. The integral over $\sigma_{0j}$  describes a \emph{charging} process from no interaction with the fluid particles ($\sigma_{0j}=0$) to the impurity becoming  a particle of species $i$ ($\sigma_{0j}=\sigma_{ij}$). The compressibility factor derived from the chemical-potential route is\cite{SR13}
\beq
Z^{(\mu)}(\rho)=1+\beta\sum_i x_i\mu^\ex_i(\rho)-\beta\int_0^1 dt\,\sum_i x_i\mu^\ex_i(\rho t).
\label{Zmu}
\eeq
In the particular case of additive mixtures, the excess term in Eq.\ \eqref{eq:102-1} becomes\cite{SR13}
\bal
\beta\mu_{i}^\ex(\rho)=&-\ln\left(1-\eta\right)+4\pi \rho\sum_{j}x_{j}\nn
&\times\int_{\frac{1}{2}\sigma_j}^{\sigma_{ij}}{d}\sigma_{0j}\,\sigma_{0j}^{2}y_{0j}(\sigma_{0j};\rho),
\label{eq:141}
\eal
where
\beq
\eta=\frac{\pi}{6}\rho M_3
\label{eta}
\eeq
is the packing fraction,
\beq
 M_n\equiv \sum_i x_i\sigma_i^n
\label{M_n}
\eeq
being the $n$th moment of the size distribution.

\subsection{Virial expansion}
In the low-density regime, the compressibility factor can be represented as a series expansion in powers of density,
\beq
Z(\rho)=1+B_2\rho+B_3\rho^2+B_4\rho^3+\cdots.
\label{v_exp}
\eeq
More generally, the radial distribution functions can also be expanded as
\beq
g_{ij}(r;\rho)=\Theta(r-\sigma_{ij})\left[1+y_{ij}^{(1)}(r)\rho+y_{ij}^{(2)}(r)\rho^2+\cdots\right],
\label{gij}
\eeq
with
\beq
y_{ij}^{(1)}(r)=\sum_k  x_k \mathcal{V}_{\sigma_{ik},\sigma_{kj}}(r),
\label{y1ij}
\eeq
where $\mathcal{V}_{a,b}(r)$ is the intersection volume of two spheres of radii $a$ and $b$ whose centers are a distance $r$ apart.{\cite{SG96}} Its expression can be found in Appendix \ref{appA}.
Equations \eqref{v_exp}--\eqref{y1ij} hold both for additive and nonadditive HS mixtures. Henceforth, however, we will specialize to the additive case.

Insertion of Eq.\ \eqref{gij} into Eq.\ \eqref{hij} yields
\beq
\hat{h}_{ij}(\rho)=\sqrt{x_ix_j}\rho\left[-\frac{4\pi}{3}\sigma_{ij}^3+H_{ij}^{(1)}\rho+H_{ij}^{(2)}\rho^2+\cdots\right],
\eeq
with
\bal
H_{ij}^{(1)}=&4\pi\int_{\sigma_{ij}}^\infty dr\, r^2y_{ij}^{(1)}(r)\nn
=&\frac{\pi^2}{36}\Big[M_6+6(M_5+2M_3\sigma_i\sigma_j)\sigma_{ij}
+
3M_4(4\sigma_{ij}^2+\sigma_i\sigma_j)\Big],
 \label{H1ij}
\eal
\beq
H_{ij}^{(2)}=4\pi\int_{\sigma_{ij}}^\infty dr\, r^2y_{ij}^{(2)}(r).
\label{H2ij}
\eeq
{}From Eqs.\ \eqref{eq:104} and \eqref{Zc} it is easy to obtain
\bal
B_2&=\frac{2\pi}{3}\sum_{i,j}x_ix_j\sigma_{ij}^3\nn
&=\frac{\pi}{6}\left(3M_1M_2+M_3\right),
\label{B2c}
\eal
\bal
B_3&=\frac{16\pi^2}{27}\sum_{i,j,k}x_ix_jx_k\sigma_{ik}^3\sigma_{kj}^3-\frac{1}{3}\sum_{i,j}x_ix_jH_{ij}^{(1)}\nn
&=\left(\frac{\pi}{6}\right)^2\left(6M_1M_2M_3+3M_2^3+M_3^2\right),
\label{B3c}
\eal
\bal
B_4^{(c)}=&\frac{16\pi^3}{27}\sum_{i,j,k,\ell}x_ix_jx_kx_\ell\sigma_{ik}^3\sigma_{k\ell}^3\sigma_{\ell j}^3\nn
&
-\frac{2\pi}{3}\sum_{i,j,k}x_ix_jx_k\sigma_{ik}^3H_{kj}^{(1)}-\frac{1}{4}\sum_{i,j}x_ix_jH_{ij}^{(2)}.
\label{B4c}
\eal

The virial route is simpler. For additive mixtures, Eq.\ \eqref{y1ij} implies
\beq
y_{ij}^{(1)}(\sigma_{ij})=\frac{\pi}{6}\left(M_3+\frac{3}{2}M_2\frac{\sigma_i\sigma_j}{\sigma_{ij}}\right).
\label{y1ijcont}
\eeq
Thus, from Eq.\ \eqref{eq:101-1} one recovers Eqs.\ \eqref{B2c} and \eqref{B3c}. As for the fourth virial coefficient, Eq.\ \eqref{eq:101-1} gives
\beq
B_4^{(v)}=\frac{2\pi}{3}\sum_{i,j}x_ix_j\sigma_{ij}^3y_{ij}^{(2)}(\sigma_{ij}).
\label{B4v}
\eeq
The fourth virial coefficients \eqref{B4c} and \eqref{B4v} are not expected to agree unless the exact functions $y_{ij}^{(2)}(r)$ are used.

As for the chemical-potential route,  we note that $y_{0j}(\sigma_{0j};\rho)$ is given by the terms inside the square brackets in Eq.\ \eqref{gij}. In particular, the exact quantity $y_{0j}^{(1)}(\sigma_{0j})$ is  obtained from Eq.\ \eqref{y1ijcont} by the changes $\sigma_{ij}\to\sigma_{0j}$ and $\sigma_i\to 2\sigma_{0j}-\sigma_j$. As a consequence, Eq.\ \eqref{eq:141} becomes
\begin{widetext}
\bal
\beta\mu_i^\ex=&\frac{\pi \rho}{6}\left(M_3+3M_2\sigma_i+3M_1\sigma_i^2+\sigma_i^3\right)
+\left(\frac{\pi \rho}{6}\right)^2\Big[\frac{M_3^2}{2}+3M_2M_3\sigma_i+\left(3M_1M_3+\frac{9M_2^2}{2}\right)\sigma_i^2\nn
&+\left(3M_1M_2+M_3\right)\sigma_i^3\Big]
+\left(\frac{\pi\rho}{6}\right)^3\left[\frac{M_3^3}{3}+\frac{864}{\pi^2}
\sum_{j}x_{j}\int_{\frac{1}{2}\sigma_j}^{\sigma_{ij}}{d}\sigma_{0j}\,\sigma_{0j}^{2}y_{0j}^{(2)}(\sigma_{0j})\right]
+\cdots,
\label{muiexp}
\eal
\end{widetext}
Next, application of Eq.\ \eqref{Zmu} gives Eqs.\ \eqref{B2c}, \eqref{B3c}, and
\beq
B_4^{(\mu)}=\left(\frac{\pi}{6}\right)^3\left[\frac{M_3^3}{4}+\frac{648}{\pi^2}
\sum_{j}x_{j}\int_{\frac{1}{2}\sigma_j}^{\sigma_{ij}}{d}\sigma_{0j}\,\sigma_{0j}^{2}y_{0j}^{(2)}(\sigma_{0j})\right].
\label{B4mu}
\eeq

\subsection{Composition-independent coefficients}
The fourth virial coefficient is a fourth-degree polynomial in the mole fractions, i.e.,
\beq
B_4=\sum_{i,j,k,\ell}x_ix_jx_kx_\ell B_{ijk\ell},
\label{3a}
\eeq
where $B_{ijk\ell}$ are \emph{composition-independent} coefficients that otherwise are functions of the diameters $\sigma_i$, $\sigma_j$, $\sigma_k$, and $\sigma_\ell$.
Obviously, $B_{iiii}=(\pi/6)^3\sigma_i^9 b_4$, where
$b_4$ is the (reduced) virial coefficient of the pure system. Thus, in the case of a binary mixture ($\mm=2$) the nontrivial coefficients are $B_{1112}$, $B_{1122}$, and $B_{1222}$. In dimensionless form,
\beq
B_{1112}^*(q)\equiv\left(\frac{6}{\pi}\right)^3\sigma_1^{-9}{B}_{1112}(\sigma_1,\sigma_2),
\label{B1112}
\eeq
\beq
B_{1122}^*(q)\equiv\left(\frac{6}{\pi}\right)^3 \sigma_1^{-6}\sigma_2^{-3}{B}_{1122}(\sigma_1,\sigma_2),
\label{B1122}
\eeq
\beq
B_{1222}^*(q)\equiv\left(\frac{6}{\pi}\right)^3 \sigma_1^{-3}\sigma_2^{-6}{B}_{1222}(\sigma_1,\sigma_2),
\label{B1222}
\eeq
where $q\equiv\sigma_2/\sigma_1$ is the size ratio.
By symmetry, $B_{1222}^*(q)=q^3B_{1112}^*(1/q)$. Moreover, the reduced coefficients must satisfy a number of consistency conditions.\cite{W98b,W99,BS08,LK09} In particular,
\bal
\left.\frac{\partial B_{1112}^*(q)}{\partial q}\right|_{q=1}&=\frac{3}{2}\left.\frac{\partial B_{1122}^*(q)}{\partial q}\right|_{q=1}=3\left.\frac{\partial B_{1222}^*(q)}{\partial q}\right|_{q=1}\nn
&=\frac{9}{4}b_4,
\label{1-2-3}
\eal
\bal
\left.\frac{\partial B_{1112}^*(q)}{\partial q}\right|_{q=0}&=\frac{9}{4},\nn
\lim_{q\to 0}B_{1112}^*(q)&=\frac{1}{4},\quad \lim_{q\to 0}B_{1122}^*(q)=2,
\label{7-4-5}
\eal
\beq
\lim_{q\to 0}B_{1222}^*(q)=\frac{15}{2}.
\label{6}
\eeq

\subsection{Exact results. Analytical and semi-empirical expressions for $B_{ijk\ell}^*$}

The exact form of the cavity functions to second order in density, $y_{ij}^{(2)}(r)$, is not known. As a consequence, the exact  composition-independent fourth virial coefficients $B_{ijk\ell}$ for any number of components is not known either.
However, an  interesting result refers to the case where the smallest sphere fits in the inner hole made by the other three spheres being tangent.\cite{B98} More specifically, $B_{ijk\ell}$ is analytically known if
\beq
\sigma_\ell\leq \frac{\sigma_i\sigma_j\sigma_k}{\sigma_i\sigma_j+\sigma_i\sigma_k+\sigma_j\sigma_k+
2\sqrt{\sigma_i\sigma_j\sigma_k(\sigma_i+\sigma_j+\sigma_k)}}.
\label{5a}
\eeq

In particular, in the case of binary mixtures ($\mm=2$) with $q\equiv \sigma_2/\sigma_1 <2/\sqrt{3}-1\simeq 0.1547$,\cite{B98}
\bal
B_{1112}^*(q)=&\frac{1}{4}+\frac{9q}{4}+9q^2
+\frac{21q^3}{4}+\frac{27q^4}{8}+\frac{27q^5}{40}
\nn
&
-\frac{27q^6}{5}-\frac{162q^7}{35}-\frac{81q^8}{56}-\frac{9q^9}{56}.
\label{6a}
\eal
If  $q >2/\sqrt{3}-1$, $B_{1112}^*$ is given by the right-hand side of Eq.\ \eqref{6a} plus the contribution\cite{LK09,U11}
\bal
\Delta B_{1112}^*(q)=&\frac{1}{280\pi}\Big[\frac{Q}{12}\left(10Q^6-
51Q^4+210Q^2\right.
\nn
&\left.+6976\right)-486P_1(Q^2+9)+\frac{q+1}{3} P_2\nn
&\times \left(5Q^8-28Q^6+129Q^4-124Q^2\right.
\nn&
\left.
+11\,378\right)
\Big],
\label{DeltaB4}
\eal
where $Q\equiv\sqrt{3q^2+6q-1}$, $P_1\equiv \tan^{-1} Q$, and $P_2\equiv \tan^{-1}\left[Q/(q+1)\right]$.
Because of the symmetry condition $B_{1222}^*(q)=q^3B_{1112}^*(1/q)$, the only remaining coefficient in a binary mixture is $B_{1122}^*$. It is made of six partial contributions, five of which are analytically known, while the sixth one needs to be numerically evaluated.\cite{LK09} An accurate semi-empirical approximation of the latter partial contribution was obtained by Lab\'ik and Kolafa.\cite{LK09}
The combined result is
\bal
B_{1122}^*(q)=&q^{3/2}\Big[3
\left(\sqrt{q}+\frac{1}{\sqrt{q}}\right)\left(q+\frac{11}{4}+\frac{1}{q}\right)\nn
&-\sqrt{2}\left(\frac{2}{u^3}+\frac{15}{4u}+\frac{9u}{5}\right)
-u^3 R\Big],
\label{RR}
\eal
where $u\equiv \sqrt{2/(q+q^{-1})}$ and $R$ is a rational function of $u$ whose coefficients are listed in Table II of Ref.\ \onlinecite{LK09}.

It is easy to check that Eqs.\ \eqref{6a}--\eqref{RR} are consistent with the one-component value
\beq
b_4=\frac{219\sqrt{2}-712\pi+4131\tan^{-1}\sqrt{2}}{35\pi}\simeq 18.3648
\label{4a}
\eeq
and with conditions \eqref{1-2-3}--\eqref{6}.

\section{Percus--Yevick and hypernetted-chain approximations}
\label{sec3}
In this section the fourth virial coefficient in the PY and HNC approximations from the three routes \eqref{B4c}, \eqref{B4v}, and \eqref{B4mu} are evaluated. To that end we need the approximate corresponding approximate expressions for $y_{ij}^{(2)}(r)$.

\subsection{PY}
The exact solution of the PY equation for additive HS mixtures is known for any density in Laplace space.\cite{L64,YSH98} {}From such a solution one can get $y_{ij}^{(2,\py)}(r)$. Its explicit expression is given in Appendix \ref{appA}.
In particular, the contact values are
\beq
y_{ij}^{\left(2,\py\right)}\left(\sigma_{ij}\right)=\left(\frac{\pi}{6}\right)^2M_{3}\left(M_{3}+3M_{2}\frac{\sigma_{i}\sigma_{j}}{\sigma_{ij}}\right).
\label{eq:297}
\end{equation}
Also, the integral \eqref{H2ij} becomes
\bal
H_{ij}^{(2,\py)}=&-\frac{\pi^3}{36}\Big[M_4M_5+4M_3M_4(\sigma_{ij}^2+\sigma_i\sigma_j)\nn
&+(3M_4^2+2M_3M_5+4M_3^2\sigma_i\sigma_j)\sigma_{ij}
\Big].
\eal
Inserting these results unto Eqs.\ \eqref{B4c}, \eqref{B4v}, and \eqref{B4mu} one obtains $B_4^{(\py\text{-}c)}$, $B_4^{(\py\text{-}v)}$, and $B_4^{(\py\text{-}\mu)}$, respectively. The three coefficients have the common structure
\beq
{B}_{4}=\left(\frac{\pi}{6}\right)^3M_3\left[C_1M_1M_2M_3+C_2M_2^3+C_3M_3^2\right].
\label{B4g}
\eeq
The corresponding  values of the PY numerical coefficients $C_1$, $C_2$, and $C_3$ are given in  Table \ref{tab}. Note that only the coefficient $C_2$ depends on the route.

\begin{table}
\caption{Numerical coefficients $C_1$, $C_2$, and $C_3$  in the expression of the  fourth virial coefficient [see Eq.\ \protect\eqref{B4g}] according to different routes in the PY and HNC approximations. The value $b_4=C_1+C_2+C_3$ is the (reduced) fourth virial coefficient in the one-component case.
The three last columns indicate whether the exact consistency conditions \protect\eqref{1-2-3}--\protect\eqref{6} are verified or not.
\label{tab}}
\begin{ruledtabular}
\begin{tabular} {lccccccc}
Approximation&$C_1$&$C_2$&$C_3$&$b_4$&\protect\eqref{1-2-3}&\protect\eqref{7-4-5}&\protect\eqref{6}\\
\hline
PY-$v$&$9$&$6$&$1$&$16$&Yes&Yes&No\\
PY-$\mu$&$9$&$\frac{27}{4}$&$1$&$\frac{57}{4}$&Yes&Yes&No\\
PY-$c$&$9$ &$9$ &$1$ &$19$&Yes&Yes&Yes\\
HNC-$v$&$\frac{27}{2}$&$\frac{27}{2}$&$\frac{3}{2}$&$\frac{57}{2}$&Yes&No&No\\
HNC-$\mu$&$\frac{27}{2}$&$\frac{27}{2}$&$\frac{11}{8}$&$\frac{227}{8}$&Yes&No&No\\
 HNC-$c$&$\cdots$ &$\cdots$ &$\cdots$ &$\frac{5623}{420}$&Yes&No&No\\
\end{tabular}
\end{ruledtabular}
\end{table}

Equation \eqref{B4g} applies to any number of components $\mm$. The composition-independent virial coefficients $B_{ijk\ell}$ defined by Eq.\ \eqref{3a} can be easily identified. In the particular case of a binary mixture [cf.\ Eqs.\ \eqref{B1112}--\eqref{B1222}],
\beq
B_{1112}^*(q)=\frac{C_1}{4}q(1+q+2q^2)
+\frac{C_2}{4}q^2(3+q)
+\frac{C_3}{4}(1+3q^3),
\label{B1112b}
\eeq
\bal
B_{1122}^*(q)=&\frac{C_1}{6}(1+q)(1+q+q^2)
+\frac{C_2}{2}q(1+q)
\nn&
+\frac{C_3}{2}(1+q^3),
\label{B1122b}
\eal
\beq
B_{1222}^*(q)=\frac{C_1}{4}(2+q+q^2)
+\frac{C_2}{4}(1+3q)+\frac{C_3}{4}(3+q^3).
\label{B1222b}
\eeq
{}From Eqs.\ \eqref{B1112b}--\eqref{B1222b} we can see that conditions \eqref{1-2-3} are automatically satisfied regardless of the numerical values of the coefficients $C_1$, $C_2$, and $C_3$.
On the other hand, the three conditions in \eqref{7-4-5} are fulfilled only if $C_1=9$, $C_3=1$ and $C_1+3C_3=12$, respectively. Thus, the three PY routes turn out to be consistent with Eqs.\ \eqref{1-2-3} and \eqref{7-4-5}.
As for Eq.\ \eqref{6}, it requires $2C_1+C_2+3C_3=30$, this condition being satisfied  by the compressibility route only.

\subsection{HNC}
In the case of the HNC approximation one has\cite{HM06,S13b}
\beq
y_{ij}^{(2,\hnc)}(r)=y_{ij}^{(2,\py)}(r)+\frac{1}{2}\left[y_{ij}^{(1)}(r)\right]^2.
\eeq
Consequently,
\bal
y_{ij}^{(2,\hnc)}(\sigma_{ij})=&\left(\frac{\pi}{6}\right)^2\Big[\frac{3}{2}M_3\left(M_3+{3}M_2\frac{\sigma_i\sigma_j}{\sigma_{ij}}\right)\nn
&+\frac{9}{8}M_2^2\left(\frac{\sigma_i\sigma_j}{\sigma_{ij}}\right)^2\Big],
\label{yij2HNC}
\eal
\beq
H_{ij}^{(2,\hnc)}=H_{ij}^{(2,\py)}+2\pi\int_{\sigma_{ij}}^\infty dr\, [ry_{ij}^{(1)}(r)]^2.
\eeq

Let us start by considering the virial and chemical-potential routes. By plugging Eq.\ \eqref{yij2HNC} into Eqs.\ \eqref{B4v} and \eqref{B4mu} one finds again results of the form \eqref{B4g}, except that the values of the coefficients $C_1$--$C_3$ differ from the PY ones. Those values are given in Table \ref{tab}. We observe that the exact relationship $B_4^{(\hnc\text{-}v)}=\frac{3}{2}B_4^{(\py\text{-}c)}$, valid for any interaction and any dimensionality,\cite{SM10} is indeed verified.
It is also interesting to remark that  $B_4^{(\hnc\text{-}v)}\simeq  B_4^{(\hnc\text{-}\mu)}$ since both quantities differ only in the coefficient $C_3$, which is $\frac{12}{11}\simeq 1.09$ times larger in the virial route than in the chemical-potential route. As a consequence,  $b_4^{(\hnc\text{-}v)}/b_4^{(\hnc\text{-}\mu)}=\frac{228}{227}\simeq 1.004$ in the pure fluid.
In the case of a binary mixture, the composition independent coefficients $B_{ijk\ell}^{*(\hnc\text{-}v)}$ and $B_{ijk\ell}^{*(\hnc\text{-}\mu)}$ are given by Eqs.\ \eqref{B1112b}--\eqref{B1222b} with the corresponding values of $C_1$--$C_3$. Now only the consistency conditions \eqref{1-2-3} are satisfied.

Regarding the compressibility route, one has
\beq
B_4^{(\hnc\text{-}c)}=B_4^{(\py\text{-}c)}-\frac{\pi}{2}\sum_{i,j}x_i x_j \int_{\sigma_{ij}}^\infty dr\, [ry_{ij}^{(1)}(r)]^2.
\label{HNC-c}
\eeq
As discussed in Appendix \ref{appB}, the second term on the right-hand side of Eq.\ \eqref{HNC-c} prevents $B_4^{(\hnc\text{-}c)}$ from accommodating to the simple structure of Eq.\ \eqref{B4g}. First, moments of order higher than $M_3$ are involved. Second, there exist terms that cannot be accounted for by moments since those extra terms depend on the size order of the species and thus they are not invariant under a relabeling of species. After some algebra, the obtained result is
\begin{widetext}
\bal
\left(\frac{6}{\pi}\right)^3 {B}_{4}^{(\hnc\text{-}c)}=&\left(\frac{6}{\pi}\right)^3{B}_{4}^{(\py\text{-}c)}+M_1\left(\frac{27}{40}M_1^2 M_6+\frac{63}{40}M_1M_2M_5-\frac{9}{8}M_1 M_3 M_4-\frac{9}{280}M_1M_7\right.\nn
&\left.-\frac{9}{8}M_2^2M_4-\frac{3}{2}M_2 M_3^2+\frac{3}{20}M_2M_6-\frac{3}{4}M_3M_5-\frac{3}{28}M_8\right)-M_2\left(\frac{9}{4}M_2^2M_3\right.\nn
&\left.+\frac{9}{40}M_2M_5
 +\frac{3}{4}M_3M_4+\frac{3}{280}M_7\right)-\frac{1}{8}M_3M_6-\frac{1}{84}M_9-\mathcal{B},
 \label{HNC-c2}
 \eal
 \end{widetext}
where, in the particular case of a binary mixture (assuming $q\equiv\sigma_2/ \sigma_1\leq 1$), the expression for the extra term $\mathcal{B}$ is
\bal
\mathcal{B}=&\sigma_1^9 x_1 x_2\frac{(1-q)^5}{105}\Big\{\frac{1-q}{4}\left[x_1^2(1039+393q+75q^2+5q^3)\right.\nn
&\left.
-x_2^2(1039q^3+393q^2+75q+5)\right]\nn
&-\frac{179  M_1 M_3+174 M_2^2+25 M_4}{\sigma_1^4}\Big\}.
\label{HNC-c3}
\eal

\begin{figure}[h!]
\includegraphics[width=7.5cm]{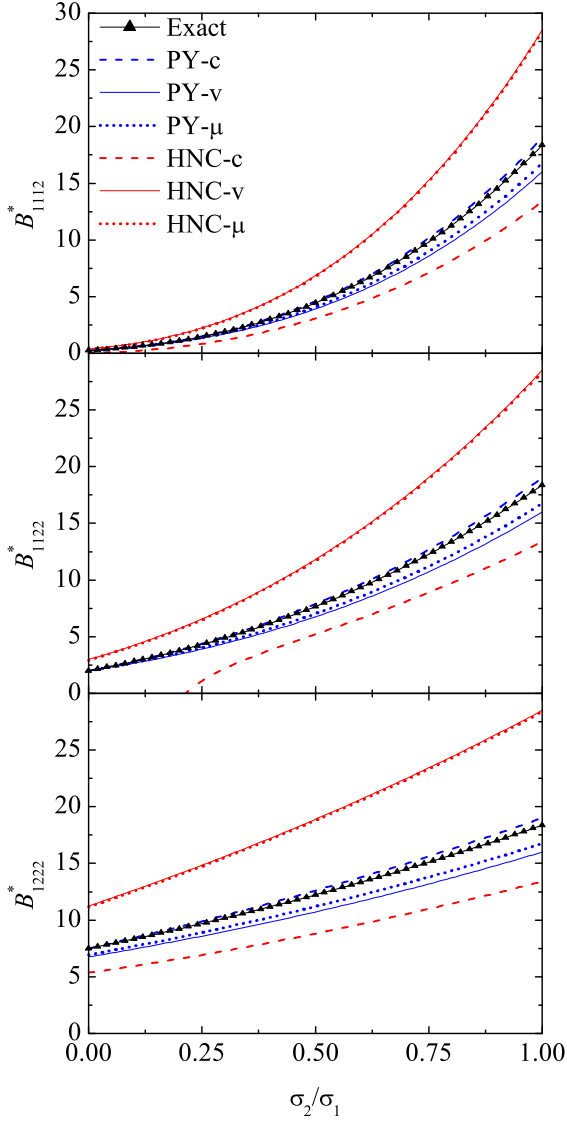}
\caption{Comparison between exact and  approximate composition-independent virial coefficients $B_{1112}^*$, $B_{1122}^*$, and $B_{1222}^*$.}
\label{fig1}
\end{figure}

In the case of the HNC compressibility  route for a binary mixture, Eqs.\ \eqref{HNC-c2} and \eqref{HNC-c3} yield (assuming again $q\leq 1$)
\bal
{B}_{1112}^{*(\text{HNC-}c)}=&-\frac{1}{1680}+\frac{621q}{560}+\frac{531q^2}{70}
+\frac{45q^3}{8}-\frac{3q^4}{2} \nn
&+\frac{3q^5}{10} +\frac{7q^6}{20}-\frac{3q^7}{140}-\frac{3q^8}{56}-\frac{q^9}{168},
\label{B1112HNC}
\eal
\bal
{B}_{1122}^{*(\text{HNC-}c)}=&-\frac{17q^{-3}}{2520}-\frac{17q^{-2}}{280}-\frac{17q^{-1}}{70}
+\frac{8}{5}+\frac{261 q}{40}\nn
&+\frac{141 q^2}{40}+\frac{5q^3}{3}+\frac{7q^4}{10}
-\frac{13q^5}{56}-\frac{31q^6}{360},
\label{B1122HNC}
\eal
\beq
{B}_{1222}^{*(\text{HNC-}c)}=\frac{43}{8}+\frac{759q}{140}+\frac{1767q^2}{560}-\frac{947q^3}{1680}.
\label{B1222HNC}
\eeq
As happened with the virial and chemical-potential routes,  only Eq.\ \eqref{1-2-3} is satisfied by ${B}_{ijk\ell}^{*(\text{HNC-}c)}$.

\section{Discussion}
\label{sec4}
Figure \ref{fig1} compares the PY and HNC predictions via the virial, compressibility, and chemical-potential routes with the exact results for the composition-independent fourth virial coefficients. Several comments are in order in view of of Fig.\ \ref{fig1} and of the results derived in Sec.\ \ref{sec3}:
\begin{enumerate}[i]
\item
As anticipated from the coefficients in Table \ref{tab}, we observe that $B_{ijk\ell}^{*(\hnc\text{-}v)}$ and $B_{ijk\ell}^{*(\hnc\text{-}\mu)}$ are practically indistinguishable.

\item
While in the PY approximation the virial and chemical-potential routes underestimate the virial coefficients and the compressibility route overestimates them, the opposite behavior is observed in the case of the HNC approximation.

\item
As mentioned above, and as a test of the results, the general property $B_{ijk\ell}^{*(\hnc\text{-}v)}=\frac{3}{2}B_{ijk\ell}^{*(\py\text{-}c)}$ is satisfied.

\item
The three PY predictions are more accurate than any of the HNC predictions.

\item
In both approximations, the chemical-potential route is slightly better than the virial one.

\item
In both approximations, the compressibility route is the most accurate one, except in the case of the HNC cofficient $B_{1122}^*$ for $\sigma_2/\sigma_1\lesssim 0.315$.

\item
The coefficients $B_{1112}^{*(\hnc\text{-}c)}$ and $B_{1122}^{*(\hnc\text{-}c)}$ become negative for $\sigma_2/\sigma_1\lesssim 0.00053$ and $\sigma_2/\sigma_1\lesssim 0.213$, respectively. Moreover, $B_{1122}^{*(\hnc\text{-}c)}$ diverges to $-\infty$ in the limit $\sigma_2/\sigma_1\to 0$.

\item
Except for $B_{4}^{*(\hnc\text{-}c)}$, the other five theoretical predictions depend on the size composition only through the first three moments and have the common structure \eqref{B4g}.

\end{enumerate}

\section{Concluding remarks}
\label{sec5}
The results presented in this paper confirm that, even though the HNC approximation retains more diagrams than the PY one,\cite{HM06,S13b} it is certainly less reliable than the latter when applied to HS systems. It is well known that the energy and virial routes are fully equivalent for any system in the HNC theory.\cite{BH76,M60} Our results show that, in addition, the chemical-potential and virial routes are practically identical, at least at the level of the fourth virial coefficient for polydisperse HS fluids. Thus, we are in the presence of a neat example showing that a high degree of internal consistency does not necessarily correlate with accuracy.

It is also interesting to note that in both theories the compressibility route (which needs the whole spatial dependence of the pair correlation functions) is generally more efficient than the virial and chemical-potential routes (which only need the contact values) in concealing the deficiencies associated with the approximate nature of the theory. On the other hand, this feature seems to be restricted to highly repulsive interactions since the addition of an attractive part (as in the sticky-hard-sphere model) tends to worsen the quality of the compressibility route and makes the chemical-potential route the most accurate one.\cite{RS14}

To conclude, we hope that this paper can contribute to a better understanding of the merits, shortcomings, and peculiarities of the two classical integral equations when applied to such an important model as the HS multicomponent fluid.

\begin{acknowledgments}
A.S.  acknowledges the financial support of the Spanish Government through Grant No.\ FIS2010-16587 and  the Junta de Extremadura (Spain) through Grant No.\ GR10158 (partially financed by FEDER funds).
\end{acknowledgments}

\appendix
\section{Supplementary equations}
\label{appA}
In this appendix we include some equations that, for conciseness, are omitted in the main text.

First, the formula for the overlap volume $\mathcal{V}_{a,b}(r)$ is
\beq
\mathcal{V}_{a,b}(r)=\begin{cases}
\frac{4\pi}{3}\min(a^3,b^3),&0<r<|a-b|,\\
W_{a,b}(r)  ,&|a-b|<r<a+b,\\
0,&r>a+b,
\end{cases}
\label{y1ijk}
\eeq
with
\beq
W_{a,b}(r)=\frac{\pi (a+b-r)^2[r^2+2(a+b)r-3(a-b)^2]}{12r}.
\label{Wab}
\eeq
{Note that the case $0<r<|a-b|$ was not considered in the Appendix of Ref.\ \onlinecite{SG96}. Moreover,  Eq.\ \eqref{Wab} is more compact than the expressions found in Ref.\ \onlinecite{SG96}.}

Next, the PY expression for $y_{ij}^{(2,\py)}(r)$ is
\begin{widetext}
\bal
y_{ij}^{(2,\py)}(r)=&\frac{\pi}{3}M_3 y_{ij}^{(1)}(r)+\left(\frac{\pi}{6}\right)^2\frac{1}{r}\Big[\sum_k x_k \Theta(\sigma_{k}+\sigma_{ij}-r) (\sigma_{k}+\sigma_{ij}-r)^2F_{ij;k}(r) \nn
&+\sum_{k,\ell} x_k x_\ell\Theta(\sigma_{k}+\sigma_{\ell}+\sigma_{i j}-r)  (\sigma_{k}+\sigma_{\ell}+\sigma_{i j}-r)^4F_{ij;k\ell}(r)\Big],
\eal
where
  \bal
F_{ij;k}(r)=&\frac{(\sigma_k+\sigma_{ij}-r)^2}{70}\Big[4(\sigma_{ij}-r)^2(6\sigma_{ij}+r)
+21(\sigma_i\sigma_j-M_2)(3\sigma_k+2r-2\sigma_{ij})+
\sigma_k\sigma_{ij}\nn
&\times(9\sigma_k+38r-54\sigma_{ij})-2\sigma_k(\sigma_k+r)(9\sigma_k-8r)\Big]
+M_1\frac{3(\sigma_k+\sigma_{ij}-r)}{10}\nn
&\times\Big\{5\sigma_i\sigma_j(3\sigma_k+r-\sigma_{ij})
+2\sigma_{ij}(r-\sigma_{ij})\left(8\sigma_k+r-\sigma_{ij}\right)
+
\sigma_k[2(r-\sigma_{ij})^2\nn
&-\sigma_k(3\sigma_k-\sigma_{ij}-r)]
\Big\}
+M_2\Big\{9\sigma_i\sigma_j\sigma_k+\frac{3}{2}
\sigma_{ij}\left[\sigma_k^2+6\sigma_k(r-\sigma_{ij})-(r-\sigma_{ij})^2\right]\Big\},
\eal
\bal
F_{ij;k\ell}(r)=&-\frac{6}{35}(\sigma_{k}+\sigma_{\ell}+\sigma_{i j}-r)^3-\frac{r}{5}(\sigma_{k}+\sigma_{\ell}+\sigma_{i j}-r)^2+\frac{6}{5}(\sigma_{k}+\sigma_{\ell}+\sigma_{i j}-r)\nn
&\times (\sigma_{ik}\sigma_{\ell j}+\sigma_{ik}\sigma_{k\ell}+\sigma_{k\ell}\sigma_{\ell j})-6\sigma_{ik}\sigma_{k\ell}\sigma_{\ell j}.
\eal
\end{widetext}

\section{The compressibility route in the HNC approximation}
\label{appB}
In the application of the compressibility route  in the HNC approximation [see Eq.\ \eqref{HNC-c}] one has to deal with the term
\beq
\sum_{i,j}x_i x_j\int_{\sigma_{ij}}^\infty  dr\, [r y_{ij}^{(1)}(r)]^2.
\label{B1}
\eeq
According to Eqs.\ \eqref{y1ij} and \eqref{y1ijk}, the mathematical structure of $[y_{ij}^{(1)}(r)]^2$ for $r>|\sigma_i-\sigma_j|$ is
\bal
[y_{ij}^{(1)}(r)]^2=&\sum_{k,\ell}x_k x_\ell \Theta(\sigma_{ij}+\sigma_k-r)
\Theta(\sigma_{ij}+\sigma_\ell-r)\nn
&\times W_{\sigma_{ik},\sigma_{kj}}(r) W_{\sigma_{i\ell},\sigma_{\ell j}}(r)\nn
=&\sum_{k,\ell}x_k x_\ell \Theta(\sigma_{ij}+\sigma_k-r)
 W_{\sigma_{ik},\sigma_{kj}}(r) W_{\sigma_{i\ell},\sigma_{\ell j}}(r)\nn
& -\sum_{k,\ell}x_k x_\ell \Theta(\sigma_{ij}+\sigma_k-r)
\Theta(r-\sigma_{ij}-\sigma_\ell)\nn
&\times W_{\sigma_{ik},\sigma_{kj}}(r) W_{\sigma_{i\ell},\sigma_{\ell j}}(r),
\eal
where in the second step we have used the property $\Theta(x)=1-\Theta(-x)$.
Now, without loss of generality, we assume that $\sigma_1\geq\sigma_2\geq \cdots\geq \sigma_s$. In that case,
$\Theta(\sigma_{ij}+\sigma_k-r)
\Theta(r-\sigma_{ij}-\sigma_\ell)=0$ if $k\geq \ell$. Therefore,
\bal
[y_{ij}^{(1)}(r)]^2
=&\sum_{k,\ell}x_k x_\ell \Theta(\sigma_{ij}+\sigma_k-r)
 W_{\sigma_{ik},\sigma_{kj}}(r) W_{\sigma_{i\ell},\sigma_{\ell j}}(r)\nn
& -\sum_{k<\ell}x_k x_\ell \Theta(\sigma_{ij}+\sigma_k-r)
\Theta(r-\sigma_{ij}-\sigma_\ell)\nn
&\times W_{\sigma_{ik},\sigma_{kj}}(r) W_{\sigma_{i\ell},\sigma_{\ell j}}(r).
\label{B3}
\eal
When inserted into Eq.\ \eqref{B1}, the first term on the right-hand side of Eq.\ \eqref{B3} gives rise to the contribution in Eq.\ \eqref{HNC-c2} expressed in terms of the first nine moments of the size distribution. On the other hand, the contribution associated with the second term on the right-hand side of Eq.\ \eqref{B3} is not invariant under a relabeling of indices because of the constraint $k<\ell$. Such a contribution is given by Eq.\ \eqref{HNC-c3} in the particular case of a binary mixture.

\bibliographystyle{apsrev}

\bibliography{D:/Dropbox/Public/bib_files/liquid}

\end{document}